\documentclass[reprint,showpacs,superscriptaddress]{revtex4-1}
\usepackage{amsmath}
\usepackage{amssymb}
\usepackage{graphicx}
\usepackage{hyperref}
\usepackage{url}

\usepackage[resetlabels,labeled]{multibib}

\usepackage{amsfonts}
\usepackage[T1]{fontenc}
\usepackage[latin9]{inputenc}
\usepackage{multirow}
\usepackage{longtable}
\usepackage{tabu}
\usepackage{float}
\usepackage{array}
\usepackage[usenames,dvipsnames]{color}
\makeatletter
\providecommand{\tabularnewline}{\\}
\makeatother
\usepackage[english]{babel}
\usepackage{color}

\begin{document}

\title{Catalogue of Topological Electronic Materials}

\author{Tiantian Zhang}
\thanks{These authors contributed to this work equally.}
\affiliation{Beijing National Labratory for Condensed Matter Physics, and Institute
of Physics, Chinese Academy of Sciences, Beijing 100190, China}
\affiliation{University of Chinese Academy of Sciences, Beijing 100049, China}

\author{Yi Jiang}
\thanks{These authors contributed to this work equally.}
\affiliation{University of Chinese Academy of Sciences, Beijing 100049, China}
\affiliation{Beijing National Labratory for Condensed Matter Physics, and Institute
of Physics, Chinese Academy of Sciences, Beijing 100190, China}

\author{Zhida Song}
\thanks{These authors contributed to this work equally.}
\affiliation{Beijing National Labratory for Condensed Matter Physics, and Institute
of Physics, Chinese Academy of Sciences, Beijing 100190, China}
\affiliation{University of Chinese Academy of Sciences, Beijing 100049, China}

\author{He Huang}
\affiliation{Computer Network Information Center, Chinese Academy of Sciences,
China}

\author{Yuqing He}
\affiliation{Computer Network Information Center, Chinese Academy of Sciences,
China}
\affiliation{University of Chinese Academy of Sciences, Beijing 100049, China}

\author{Zhong Fang}
\affiliation{Beijing National Labratory for Condensed Matter Physics, and Institute
of Physics, Chinese Academy of Sciences, Beijing 100190, China}

\author{Hongming Weng}
\email{hmweng@iphy.ac.cn}
\affiliation{Beijing National Labratory for Condensed Matter Physics, and Institute
of Physics, Chinese Academy of Sciences, Beijing 100190, China}

\author{Chen Fang}
\email{cfang@iphy.ac.cn}
\affiliation{Beijing National Labratory for Condensed Matter Physics, and Institute
of Physics, Chinese Academy of Sciences, Beijing 100190, China}
\affiliation{CAS Centre for Excellence in Topological Quantum Computation, Beijing, China}

\maketitle
\textbf{
Topological electronic materials are new quantum states of matter hosting novel linear responses in the bulk and anomalous gapless states at the boundary, and are for scientific and applied reasons under intensive research in physics and in materials sciences\cite{Hasan2010,Qi2011,Chiu2016,Armitage2018}.
The detection for such materials has so far been hindered by the level of complication involved in the calculation of the so-called topological invariants, and is hence considered a specialized task that requires both experience with materials and expertise with advanced theoretical tools. 
Here we introduce an effective, efficient and fully automated algorithm in obtaining the topological invariants for \textit{all} non-magnetic materials that are known to human, based on recently developed principles\cite{mp-551685_mp-21861_mp-1777_mp-23224_mp-12359_mp-19976_mp-1052_mp-30050_mp-34765} that allow for exhaustive mappings between the symmetry representation of occupied bands and the topological invariants\cite{Po2017,Song2017a,Fang2017}. 
Our algorithm requires as input only the occupied-band information (energy and wavefunction) at a handful (up to eight) of high-symmetry points in the Brillouin zone, which is readily calculable with any first-principles software.
In return, it is capable of providing a detailed topological classification of \textit{all} non-magnetic materials. 
Equipped with this method we have scanned through a total of {39519} materials available in structural databases\cite{Jain2013,Ong2012,Hellenbrandt2004}, and found that as many as {8056} of them are actually topological (8889 if spin-orbital coupling is neglected). 
These are further catalogued into classes of {5005} topological semimetals\cite{zfang2003,Murakami2007}, {1814} topological insulators\cite{Kane2005,mp-1811,Fu2007} and {1237} topological crystalline insulators\cite{Fu2011}, most of which are new to human knowledge.
All the results are available and searchable at \url{http://materiae.iphy.ac.cn/}; and for each topological material, we have plotted the band structure as well as the local density of states, shown on the same website.}

\section{Context}

We begin by giving further context of the research by briefly introducing the recent development in the field. 
In the past decade, great progress has been made to understand the nontrivial topology in a band structure. 
In any band structure having direct gap at all momentum, new quantum numbers known as topological invariants can be defined in terms of the wavefunctions of all valence bands. 
Topological invariants are the defining properties of all topological materials.
The type and form of the invariants depend on the dimensionality and the symmetry of the system. 
In chronological order, time-reversal symmetry was first found to protect a $Z_{2}$ invariant in two\cite{Kane2005,mp-1811} and three dimensions\cite{Fu2007}, soon followed by crystalline symmetries such as inversion\cite{Turner2010,Hughes2011} (parity), mirror planes\cite{mp-1883_mp-19717,Chiu2013}, and others, each of niwhich brings in some new and independent topological invariant\cite{Fu2011,Alexandradinata2014,Fang2015,Shiozaki2015,Wang2016,Wieder2017,Song2017,Fang2017a,Song2017a,Khalaf2017}.  
A full characterization of the topology of a given crystal hence amounts to listing all these invariants protected by all elements in the corresponding space group {(SG)}.

Parallel to this line of progression is the emergence of the field of topological semimetals\cite{Burkov2016,Armitage2018}, where the conduction and the valence bands have band crossings, \emph{i.~e.}, topological nodes, robust against symmetry-preserving perturbations.
Depending on the degeneracy and dimensionality of the nodes, topological semimetals are further classified into nodal-point and nodal-line semimetals. 
A topological semimetal is characterized by the number and the type of all its band crossings.

Numerical prediction of topological materials thus requires the evaluation of all topological invariants or the identification of all topological nodes, both of which require lengthy and involved calculation.
The challenge have prevented people from doing a thorough scanning for topological materials, and successful examples have been mostly ascribed to experience and intuition of experienced researchers.

A series of recent theoretical works have greatly improved the situation by finding the complete mapping from the irreducible representations of valence bands to topological invariants and topological nodes\cite{mp-551685_mp-21861_mp-1777_mp-23224_mp-12359_mp-19976_mp-1052_mp-30050_mp-34765,Elcoro2017,Po2017,Song2017a,Fang2017,Khalaf2017}.
By recognizing that these theories can be fused together with first principles numerical methods, the latter of which provide the input for the former, here we develop a fully automated search algorithm that can readily be used to scan through large materials databases.

\section{The algorithm}

\begin{figure*}
\begin{centering}
\includegraphics[scale=0.35]{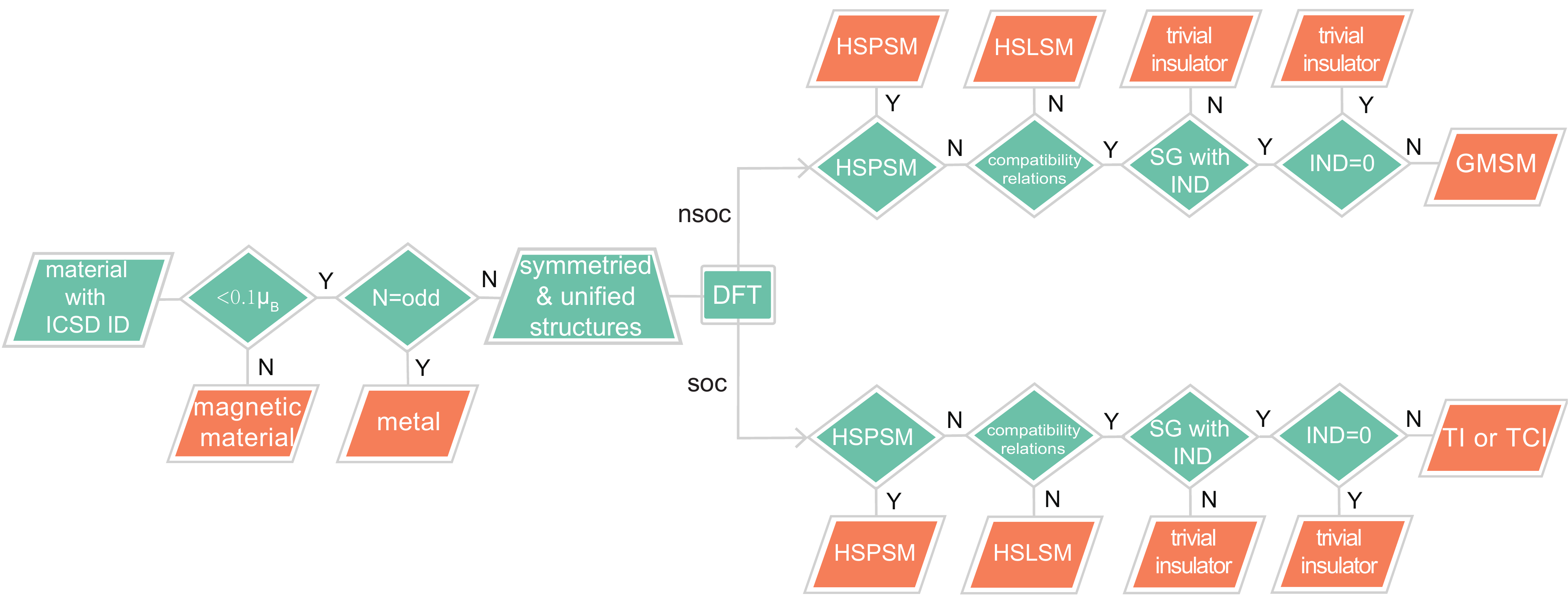} 
\par\end{centering}
\protect\caption{\label{fig:flowchart}The flowchart shows how a material is processed in our diagnosis algorithm, starting from the left. We first check against the record at the Materials Project if the material has ferromagnetism, and check if there are even number of electrons in one primitive unit cell. If yes for both, we feed the material into the first-principles calculation for the band structure and compute the symmetry data, before checking if there be partially filled irreducible representation at high-symmetry points. If not, the symmetry data is checked against all compatibility relations, and, should all relations be satisfied, is fed into the calculator for symmetry-based indicators (IND). At each checkpoint, a material either goes on to the next step, or is labeled as magnetic, conventional metal, high-symmetry point semimetal or high-symmetry line semimetal. At the final step, depending on the values of indicators, a material is labeled as generic momentum semimetal, topological insulator, topological crystalline insulator, or trivial insulator. From the first-principles calculation step, all steps are taken for two settings: non-spin-orbital-coupling setting and spin-orbital-coupling setting.}
\end{figure*}

We now describe the full diagnosis algorithm for an arbitrary crystalline material, which is summarized in a flow chart in Fig. \ref{fig:flowchart}.
For each material that is shared in both the Materials Project\cite{Jain2013,Ong2012} and the International Crystal Structure Database\cite{Hellenbrandt2004}, we load its basic information from the Materials Project.
The established mapping between symmetry data and topology data only applies to non-magnetic materials, therefore as first step we check against record if the total magnetization in one unit cell is less than $10^{-1}\mu_{B}$.
If not, we label it as ``magnetic'' and load the next material; if yes, we proceed to the second step. 
In the second step we check if the number of electrons in one unit cell is even.
If it is odd, this material is labeled as ``conventional metal'', but if even, we proceed to loading its structural data and atomic positions from the Materials Project.
The third step is symmetrizing and standardizing the atomic positions using PHONOPY\cite{Togo2015}\footnote{166 materials are discarded at this step for discrepancy in the space groups identified in PHONOPY and those given on the databases as we decrease the error tolerance in atomic position from default $10^{-5}$ \AA \  to 0.1 \AA.}.
This concludes the preparation phase of the process, after which out of {39519} scanned materials, {10348} materials are labeled ``magnetic'' and {2483} labeled ``conventional metal'', and {26688} materials go to the next phase.

We load the atomic positions and pseudo potential information of each remaining material into Vienna Ab initio Simulation Package (VASP\cite{Kresse1993,Kresse1994,Kresse1996,Kresse1996a}, which may be replaced with any other software of choice). 
The first principles calculation is done with two settings, called ``soc-setting'' and ``nsoc-setting'' where we turn on and off the spin-orbital coupling, respectively. 
The two settings, mathematically, correspond to two symmetry classes, the symplectic and the orthogonal classes\cite{Altland1997}, the mappings between symmetry data and topology data drastically differ in these two classes\cite{Song2017,Khalaf2017,Fang2017}.
In realistic electronic materials, spin-orbital coupling is never exactly zero, yet for systems consisting of small atoms, the absence of spin-orbital coupling is a relevant approximation\cite{Burkov2011,Weng2015,Fang2015a}.
For materials with high atomic numbers and high orbital electrons, where large spin-orbital coupling exists, the results obtained in the nsoc-setting should be used with caution. 
The output of the calculation is a list of energy levels and corresponding Bloch wavefunctions at each high-symmetry point. 
A high-symmetry point (HSP) is a momentum in the Brillouin zone, the invariant subgroup of which is larger than that of any point in its neighborhood. 
For example, $\Gamma$ point is always a HSP, as it is the only point in the neighborhood that is invariant under time-reversal symmetry.
With such information collected for all {26688} materials, we proceed to the analysis phase.

In the analysis phase, we first determine the irreducible representations for each valence band, or in the case of higher than one dimensional irreducible representations, for each degenerate multiplet of valence bands. 
The valence bands are defined as the lowest $N$ bands in the band structure, where $N$ is the number of electrons per unit cell.
Here the ``irreducible representations'' include the degenerate irreducible representations that are pinned at the same energy due to time-reversal symmetry\cite{Bradley2010}.
To this end, we calculate the character for each symmetry operation, and, by comparing these characters to character tables on the Bilbao Crystallographic Server (BCS), recently available thanks to decent efforts\cite{Elcoro2017}, we identify the irreducible representation for each (multiplet of) valence band(s).
Then, we check at each HSP if the top (multiplet of) valence band(s) is partially filled.
For example, if the top valence bands belong to a four dimensional irreducible representation, and if the filling number ($N$ minus the number of bands below the top valence bands) is less than four, then we have a partially filled HSP.
Each partially filled HSP is a robust band crossing point, making the material a topological semimetal, with at least one topological node at the partially filled HSP. 
We then denote the corresponding material ``high-symmetry point semimetal'' (HSPSM). 
If none of the HSP is partially filled, we generate for the material a series of integers we call the ``symmetry data''.
Each entry in the data is the number of the appearances of a given irreducible representation in the valence bands at each HSP. 
We proceed to check if the symmetry data satisfy all ``compatibility relations''.
Compatibility relations are the necessary conditions for a band structure to have direct gap along certain high-symmetry lines in the Brillouin zone. 
These relations have been derived explicitly and available on the BCS\cite{mp-551685_mp-21861_mp-1777_mp-23224_mp-12359_mp-19976_mp-1052_mp-30050_mp-34765,Elcoro2017}.
Each group of compatibility relations correspond to a certain high-symmetry line joining two high-symmetry points; and if any one relation in the group is unsatisfied, there is, at least, one topological node along this line between the conduction and the valence bands. 
When that happens, we denote the material ``high-symmetry line semimetal'' (HSLSM) and also output the line on which the node should appear. 
For the nsoc-setting, {5508} materials are labeled as HSPSM and {3269} as HSLSM, and for the soc-setting, {2713} materials belong to HSPSM and {2292} to HSLSM.
If all compatibility relations are satisfied ({17157} materials in the nsoc-setting and {20745} in the soc-setting), the material may have direct gap along all high-symmetry lines.
For these materials, we proceed to compute the symmetry-based indicators.

The group structures of indicators are derived in Ref.{[}\onlinecite{Po2017}{]}, and their explicit expressions in Ref.{[}\onlinecite{Song2017a,Fang2017,Khalaf2017}{]}. 
The indicators for each space group, if exist, are a set of several $\mathbb{Z}_{n}$ numbers, and they roughly speaking quantify how any given symmetry data differs from that of an atomic insulator having the same crystal structure.
The symmetry indicators have the following properties: (i) if any indicator is nonzero, the material is \emph{not} an atomic insulator, \emph{i.~e.}, topologically nontrivial, (ii) two materials with different indicators are topologically distinct and (iii) the topological distinction between two materials having the same indicators cannot be diagnosed using symmetry data. 
We remark that point (iii) above means that all information on topology that may be extracted from symmetry data has been contained in the the values of indicators. 
In Ref.{[}\onlinecite{Song2017a,Khalaf2017}{]}, it is shown that each nonzero combination of indicators in the soc-setting correspond to some topological (crystalline) insulator, and that in the nsoc-setting correspond to a topological semimetal with nodes at generic momenta. 
In the soc-setting, the topological invariants corresponding to each nonzero set of indicators are found in Ref.{[}\onlinecite{Song2017a}{]}, and the topological surface states in Ref.{[}\onlinecite{Khalaf2017}{]}; in nsoc-setting, Ref.{[}\onlinecite{Fang2017}{]} gives for each nonzero set of indicators the number, type and topological charge (if any) of the topological nodes.
This concludes the final analysis phase and with it our topological diagnosis of all materials.

\section{Results}

Each material is now labeled with one of the following: HSPSM = high-symmetry point semimetal (both settings), HSLSM = high-symmetry line semimetal (both settings), GMSM = generic momenta semimetal (nsoc-setting only), TI = topological insulator (soc-setting only), TCI = topological crystalline insulator (soc-setting only), magnetic, conventional metal, and trivial insulator. 
Out of these, the first five classes are considered topological materials and listed by the respective class in Tables II, III, IV, V, VI in the Supplementary Materials.
In Table II, each material in HSPSM is shown together with HSP where partial fillings occur and the irreducible representations that are partially filled; and each material in Table III is shown together with the high-symmetry line(s) where compatibility relations are unsatisfied. 
Each material in GMSM, TI and TCI is shown together with the values of its symmetry-based indicators, two types of which deserve separate notes. 
Indicators displayed in blue in Table IV signify nodal lines that have Z$_2$-monopole topological charges\cite{Fang2015a}, which are yet to be found in electronic materials\cite{Li2017,Yao2017}. 
Indicators displayed in blue in Table V have their likelihood of corresponding to topological crystalline insulators having only one-dimensional helical edge states on the boundary but not having surface states\cite{Benalcazar2017,Song2017,Langbehn2017,Benalcazar2017a,Schindler2018a}, also known as ``high order topological insulators'' in recent studies\cite{Schindler2018}.
\begin{figure*}
\begin{centering}
\includegraphics[scale=0.7]{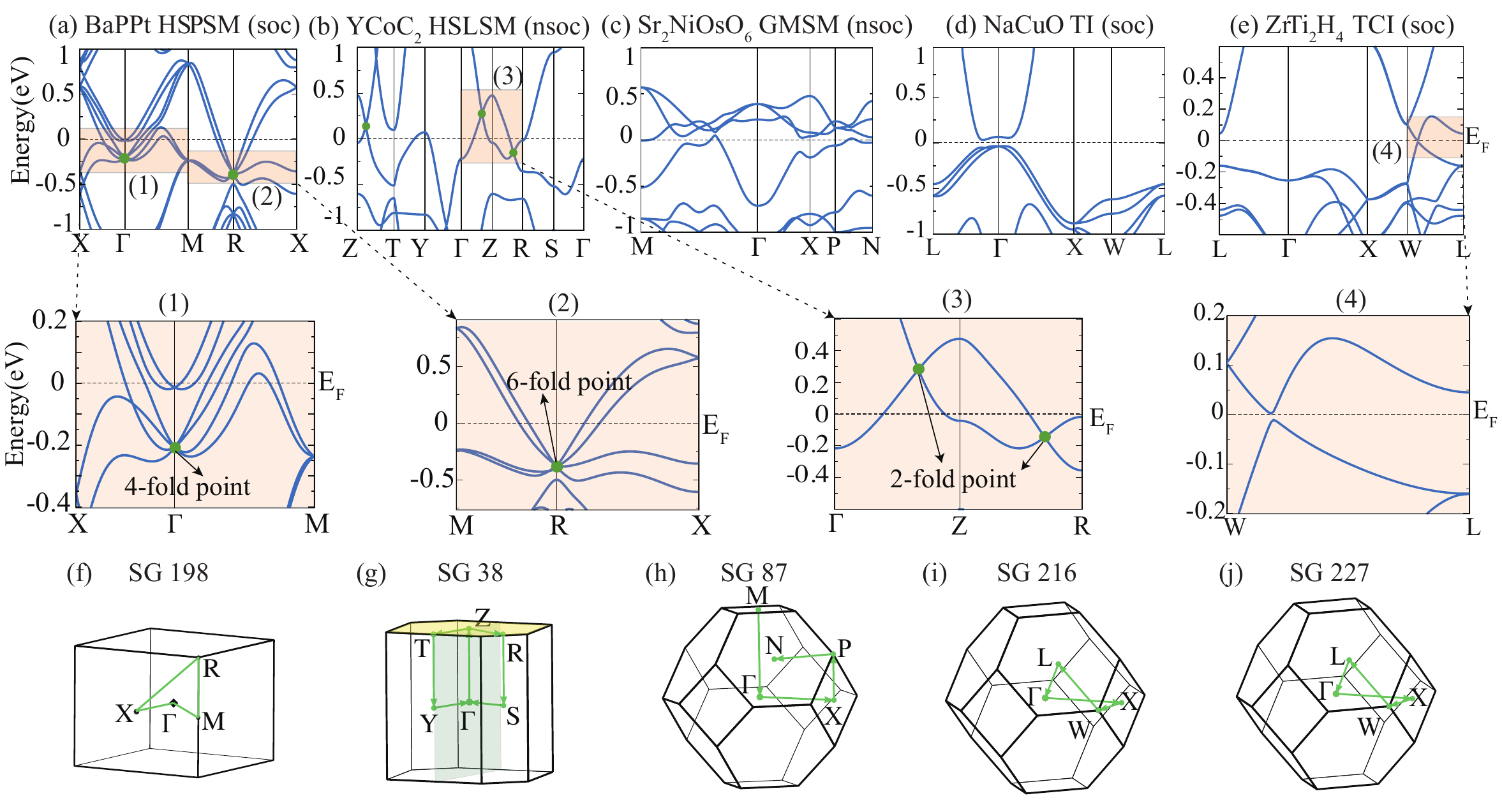} 
\par\end{centering}
\protect\caption{\label{fig:candidate}The five candidates for the five classes of topological materials: (a) high-symmetry point semimetal BaPPt having a sixfold degeneracy near the Fermi energy at R, (b) high-symmetry line semimetal YCoC$_2$ which has two connected nodal rings (nsoc-setting), (c) generic momenta semimetal Sr$_2$NiOsO$_6$ having nodal rings that have $\mathbb{Z}_2$ monopole charge (nsoc-setting), (d)  topological insulator NaCuO and (e) topological crystalline insulator ZrTi$_2$H$_4$. For each candidate material, we plot the band structure in (a-e), the Brilluoin zone with high-symmetry points marked in (f-j), and if necessary, zoomed-in regions of the band structure (1-4).}
\end{figure*}
The exhaustive scan, naturally, not only yields almost all (nonmagnetic) topological materials that have ever been so far predicted in theory or verified in experiments, but also predicts many more topological materials that have not appeared in literature.
Out of these new candidates, we pick one material in each class for display.
BaPPt in Fig.~\ref{fig:candidate}{(a)} is found a HSPSM where the conduction and the valence bands meet at $\Gamma$ and R.
The degeneracy at $R$ is sixfold stabilized by nonsymmorphic space group symmetries, and importantly, there is an electron pocket near R.
These facts qualify BaPPt as a good candidate for study of ``new fermions'' beyond Weyl and Dirac fermions in real materials\cite{Bradlyn2016}.
In the HSLSM class (nsoc-setting), we pick YCoC$_2$, the band structure of which is shown in Fig.~\ref{fig:candidate}(b).
The band structure shows that band crossings are along {Z-$\rm{\Gamma}$, Z-T, Z-Y, and Z-R} lines, and a more detailed analysis (in Sec.~\ref{sec:SII}) shows that the band crossings in this materials are nodal lines protected by mirror planes, and the nodal rings touch each other.
Sr$_2$NiOsO$_6$ (nsoc-setting) in Fig.~\ref{fig:candidate}{(c)} has no band crossing along any of the high-symmetry lines, but the indicators of $(0002)$ imply that at generic momentum there must be 2 mod 4 nodal rings where the conduction and the valence bands cross.
Each of the ring has $\mathbb{Z}_2$ topological charge, making Sr$_2$NiOsO$6$ the first candidate electronic material (with small spin-orbital coupling) hosting $\mathbb{Z}_2$ nontrivial nodal rings.
NaCuO\footnote{A related material, NaAgO, was mentioned in Ref.[\onlinecite{mp-542298_mp-568758_mp-130_mp-9027_mp-1008682_mp-1008866_mp-1009020_mp-1008807_mp-28447}].} is a new noncentrosymmetric topological insulator [Fig.~\ref{fig:candidate}(d)], featuring three band inversion between the $d$ orbital and the $s$ orbital at $\Gamma$, with the band gap $\sim0.1\ eV$.
For noncentrosymmetric systems, the classic Fu-Kane formula\cite{Fu2007} does not apply, so that an eigenvalue diagnosis would be impossible without our new method.
ZrTi$_2$H$_4$ in Fig.~\ref{fig:candidate}{(e)} has band crossings along L-W lines without spin-orbital coupling, but as the coupling turns on opens a full gap of $\sim10\ meV$ at all momenta, making the material a topological crystalline insulator.
The indicators of $(0002)$ pin down the topological invariants of the TCI to two possible sets\cite{Song2017a}.
Eigenvalue diagnosis cannot distinguish them further, but a detailed calculation (see Sec.~\ref{sec:SIII}) of the mirror Chern number at $k_z=0$-plane helps choose the correct set.
In this set, all nonzero invariants are protected by screw rotation symmetries or glide plane symmetries, so that ZrTi$_2$H$_4$ is a material candidate for a screw-axis $\mathbb{Z}_2$ TCI\cite{Song2017a,Khalaf2017}, having 1D helical edge states on its surface without 2D surface states.

All the above results are available at \url{http://materiae.iphy.ac.cn/}, which features an interactive user interface that facilitates search in the vast data. 
At the same website, we have also shown the band structures\cite{bandpath} and density of states for each material diagnosed as topological.

In summary, the exhaustive search using our new method yields a huge number of new materials predictions.
Since a candidate topological material may at the same time superconducts (such as SrSn$_4$\cite{Lin2011} and LiTi$_2$O$_4$\cite{Moshopoulou1999}), has negative coefficient of thermal expansion (such as ScF$_3$\cite{Li2011}), or possesses other unusual properties, the complete list of these materials connects the field of topological physics to many other areas of solid state physics and materials sciences.

\textit{Note added}: In the finalizing stage of the current paper, we became aware of the three recent preprints that use symmetry-based indicators for diagnosis of topological materials\cite{mp-10106,mp-542298_mp-568758_mp-130_mp-9027_mp-1008682_mp-1008866_mp-1009020_mp-1008807_mp-28447,mp-16315_mp-954_mp-570223_mp-20273_mp-614013_mp-763_mp-569568_mp-30492_mp-30875_mp-606949_mp-48}, which have covered one, five and eight out of 230 space groups in their searches, respectively.

\bibliographystyle{apsrev4-1}
\bibliography{ref}

\section{Methods}

\subsection{Setting up the first principles numerics}

All the calculations in this work are performed by Vienna $ab initio$ Simulation Package (VASP) with the generalized gradient approximation (GGA) of Perdew-Burke-Ernzerhof (PBE) type exchange-correlation potential \cite{Kresse1996a,Kresse1996,Kresse1994,Kresse1993}. The psudopotential files we used are from VASP software package and they are listed in the website \url{http://materiae.iphy.ac.cn/}. The cut-off energy of plane wave basis set is set to be the ENMAX value in the pseudo-potential file plus $25\%$. A $\Gamma$-centered Monkhorst-Pack grid with 30 k-points per $\textup{\AA}^{-1}$ is used for the self-consistent calculations\cite{MP}. {A maximum number of electronic self-consistency steps is given in our calculations, such that a material for which the calculation does not converge within 300 self-consistency loop steps is labeled and discarded. Two itinerant schemes are used in this process: the special Davidson block iteration scheme and Residual minimization method direct inversion in the iterative subspace (RMM-DIIS)\footnote{about 400 materials are not converged or converge to wrong states in the nsoc-setting and 600 in the soc-setting}.}

This diagnosis scheme is supposed to find all topological materials that \emph{may be diagnosed using symmetry eigenvalues}. 
It is emphasized that some topological materials have the \emph{same} symmetry data as atomic insulators, and so cannot be found using the scheme. 
In other words, materials not included in Tables~\ref{tab:HSPSM}, \ref{tab:HSLSM}, \ref{tab:GMSM}, \ref{tab:TI}, \ref{tab:TCI} may still be topological, but they cannot be identified without using more complicated diagnosis such as Wilson loops. 
While the mapping from symmetry data to topology data is mathematically rigorous, the validity of generalized gradient approximation depends on the actual material, so that if a material has significant correlation effect at the Fermi energy, the results are likely to be inaccurate.
{For example, for compounds containing rare-earth elements with possibly partial-filling $f$-orbitals,  the strong correlation effect is dominant and we have left them for further detailed study.} {From experience, we marked out a few elements with color that often induce strong correlation effects in compounds due to partially filled $d$- or $f$-shell in Tables~\ref{tab:HSPSM}, \ref{tab:HSLSM}, \ref{tab:GMSM}, \ref{tab:TI}, \ref{tab:TCI} as warning (blue for $d$ and red for $f$). Here Ti, Y, Zr, Nb, Mo, La, Lu, Hf, Ta, W and Pt, however, are not marked out because, while having partially filled $d$-shells, they in many known cases do not bring about strong correlation effects.}


\subsection{Proof that band crossings in YCoC$_2$ are nodal lines}
\label{sec:SII}
{Table III shows that compatibility relations for YCoC$_2$ are not satisfied along Z-$\rm{\Gamma}$, Z-T, Z-Y, and Z-R lines, where the conduction bands and valence bands touch. 
Z-$\rm{\Gamma}$ and Z-Y are on the green mirror plane, and Z-R and Z-T are on the yellow mirror plane in Fig.~\ref{fig:candidate}(g).
It is then found that the two intersecting bands have opposite mirror eigenvalues for both mirror planes, implying that the band crossings are in fact nodal lines which lying on the two mirror invariant planes.
Specially, since the band crossing point along Z-T is shared by the two mirror planes, the two nodal lines intersect each other at this point.
Therefore, there are two nodal rings, on the green and the yellow planes respectively, that are centered at Z, and the two intersect each other at two points along Z-T and minus Z-T, respectively.}


\subsection{Determine all topological invariants of {ZrTi$_2$H$_4$}}
\label{sec:SIII}
\begin{figure}
\begin{centering}
\includegraphics[scale=0.8]{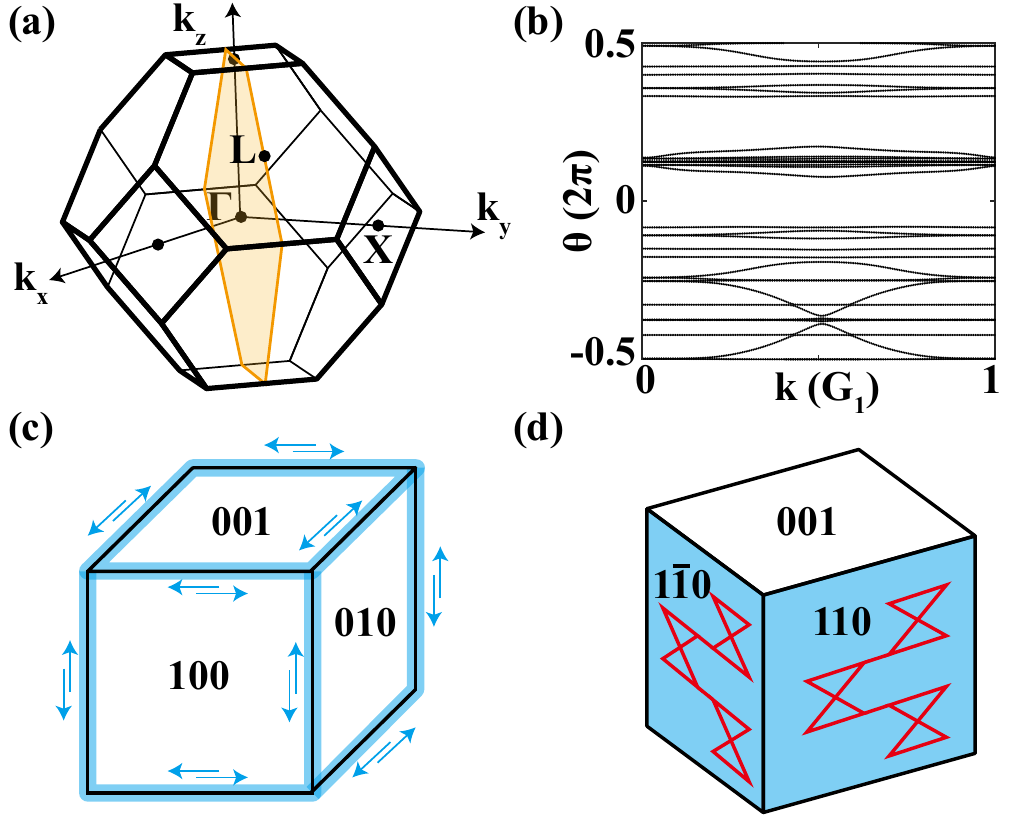} 
\par\end{centering}
\protect\caption{\label{fig:TCI} (a) Brillouin zone for ZrTi$_2$H$_4$, in which the yellow plane is  $m_{1\bar{1}0}$. (b) Wilson loop for ZrTi$_2$H$_4$ in the $m_{1\bar{1}0}$ plane. (c) One-dimensional helical modes in a cubic ZrTi$_2$H$_4$ sample. (d) Two-dimensional surface states on each surface of a cubic ZrTi$_2$H$_4$ sample.} 
\end{figure}

In order to determine the topological invariants of $\mathrm{ZrTi_{2}H_{4}}$ (space group $Fd\bar{3}m$), we look up its SI set, which is $\left(z_{2w,1},z_{2w,2},z_{2w,3},z_{4}\right)=\left(0002\right)$, in table \ref{tab:227ind} of Ref. \cite{Song2017a} and find that there are only two possibilities for the invariants for this SI set. 
In the first case, the mirror Chern number on the $1\bar{1}0$ plane in the Brillouin zone (the yellow plane in Fig. \ref{fig:TCI}) is 2 (mod 4), whereas in the second case this mirror Chern number is 0 (mod 4). 
By \textit{ab-initio} calculation, as decribed in the next paragraph, we find that this mirror Chern number is 0 and thus  $\mathrm{ZrTi_{2}H_{4}}$ belongs to the second case. 
In this case, nontrivial TCI invariants include: (d,i) hourglass invariant protected by glide plane $\left\{ m_{001}|\frac{1}{4}\frac{1}{4}0\right\} $, (ii) rotation-invariant protected by rotation $\left\{ 2_{1\bar1 0}|000\right\} $, (iii) inversion-invariant, (iv) screw-invariant protected by $\left\{ 4_{001}|0\frac{1}{4}\frac{1}{4}\right\}$, as well as those invariants protected by symmetries equivalent with above symmetries. 
All of these invariants are $\mathbb{Z}_2$ type and correspond to either two-dimensional or one-dimensional anomalous surface states. 
Here we propose two real space configurations to detect such surface states.
In Fig. \ref{fig:TCI}, we show the one-dimensional helical mode protected by screws and/or inversion.
The cubic sample is cut out along $100$, $010$, and $001$ surfaces, all of which are fully gapped. 
As long as the cubic preserves inversion symmetry there must be an inversion-symmetric one-dimensional helical mode on the boundary, whose shape depends on the experimental situation.
However, in presence of the four-fold screw symmetries, which protect nontrivial screw-invariants, the shape of helical mode is further constrained. 
We consider the sample is large enough such that the four-fold screw symmetry, $\left\{ 4_{001}|0\frac{1}{4}\frac{1}{4}\right\} $, is preserved on the side surfaces far away from the top and bottom surfaces.
Then, due to discussion in Ref. \cite{Song2017a,Khalaf2017}, four one-dimensional helical modes run along the screw axis and transform to each other in turn under the screw operation.
Similarly, along any equivalent screw axis there also {exist} four one-dimensional helical modes. 
The helical mode shown in Fig. \ref{fig:TCI} is a configuration satisfying all the above symmetry conditions.
In Fig. \ref{fig:TCI}, we show the two-dimensional surface states protected by glide and/or  two-fold rotation symmetries. 
The sample is cut out along $110$, $1\bar{1}0$, and $001$ surfaces, wherein the $001$ surface is fully gapped whereas the $1\bar{1}0$ and {$110$} surfaces are gapless. 
Due to the hourglass-invariant protected by $\left\{ m_{001}|\frac{1}{4}\frac{1}{4}0\right\}$ symmetry the $1\bar{1}0$ surface must have a hourglass mode, and due to the rotation-invariant protected by $\{2_{1\bar{1}0}|000\}$ the $1\bar{1}0$ surface must have 2 (mod 4) Dirac nodes.
In fact, the two kinds of surface states are consistent with each other: at an even filling number, which is necessary for insulator in presence of time-reversal, the two hourglass crossings protected by glide symmetry also play the role of Dirac nodes for the rotation-invariant.
Therefore the $1\bar10$ surface has a $C_2$-symmetric hourglass mode.
The $110$ surface has a similar surface state since it is equivalent with $1\bar10$ surface.

Now let us briefly describe how we calculate the mirror Chern number \textit{ab-initioly}. 
First, the parallelogram spaned by $\mathbf{G}_{1}$ and $\mathbf{G}_{2}$ shown in Fig. \ref{fig:TCI} is recognized as the minimal periodic cell in the mirror plane, wherein $\mathbf{G}_1$ is along $110$-direction and $\mathbf{G}_2$ is along $111$-direction. 
We therefore calculate the mirror Chern number within this parallelogram.
For each point along the $\Gamma\mathbf{G}_{1}$-line, $k\mathbf{G}_1$, we define a Wilson loop matrix as

\begin{align*} %
W_{nn^{\prime}}\left(k\right)  =\\
& \sum_{n_{1}\cdots n_{N-1}\in occ}\left\langle u_{n,k\mathbf{G}_{1}}\right|u_{n_{1},k\mathbf{G}_{1}+\frac{\mathbf{G}_{2}}{N}}\rangle\left\langle u_{n_{1},k\mathbf{G}_{1}+\frac{\mathbf{G}_{2}}{N}}\right|\cdots\nonumber \\
& \times\left|u_{n_{N-1},k\mathbf{G}_{1}+\left(N-1\right)\frac{\mathbf{G}_{2}}{N}}\right\rangle \left\langle u_{n_{N-1},k\mathbf{G}_{1}+\left(N-1\right)\frac{\mathbf{G}_{2}}{N}}\right| \\
&\hat{V}^{\mathbf{G}_{2}}\big|u_{n_{N-1},k\mathbf{G}_{1}}\big\rangle
\label{eq:WL} \tag{1}
\end{align*} 

where $N$ is a large enough integer to describe the infinite limit $N\to\infty$, $\left|u_{n,\mathbf{k}}\right\rangle$ the periodic part of the Bloch wave-function, $n,n^{\prime}$,$n_{i}$ ($i=1\cdots N-1$) the occupied band indices, and $\hat{V}^{\mathbf{G}_{2}}$ the embedding operator{\cite{MCN_aris}}. 
On the other hand, for each $k\mathbf{G}_1$ we can define the mirror representation matrix as
\begin{equation} \tag{2}
M_{nn^{\prime}}\left(k\right)=\left\langle u_{n,k\mathbf{G}_{1}}\right|\hat{M} \left|u_{n^{\prime},k\mathbf{G}_{1}}\right\rangle,
\end{equation}
where $\hat{M}$ is the operator of symmetry operation $\{m_{1\bar{1}0}|000\}$.
Since each $k$-point in Eq. (\ref{eq:WL}) is mirror-invariant, one can prove that $M(k)$ always commutes with $W(k)$.
Therefore, we can project the Wilson loop matrix into the subspace having mirror eigenvalue $+i$, then the mirror Chern number is just given by the winding number of the projected Wilson loop.
In Fig. \ref{fig:TCI}, we plot the eigenvalues of the projected Wilson loop matrix as a function of $k$, from which one can find that the winding number is 0.

\begin{table}
\begin{centering}
\includegraphics[scale=0.3]{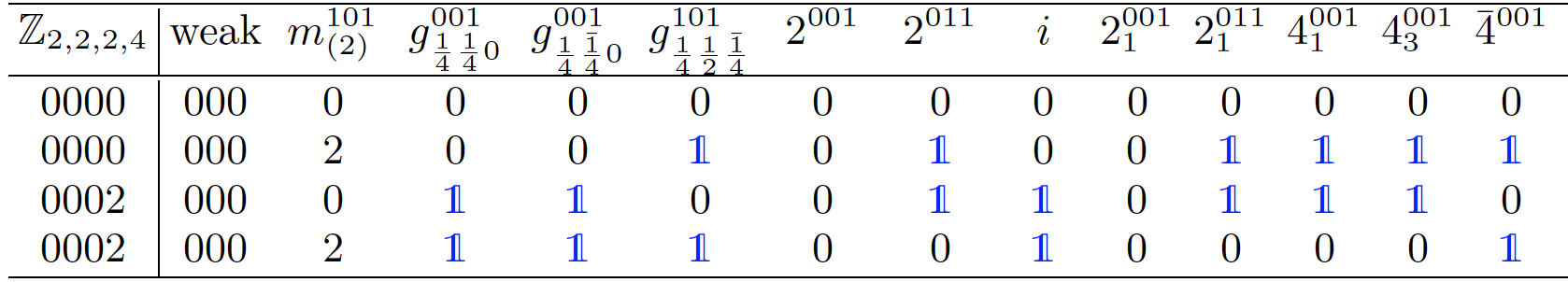} 
\par\end{centering}
\protect\caption{\label{tab:227ind} Possible invariants for SG 227.}
\end{table}

\clearpage
\newpage
\onecolumngrid
%

\twocolumngrid

\clearpage
\newpage

\setcounter{table}{1}
\clearpage
\newpage
\onecolumngrid
\section*{Supplementary Materials: tables of topological materials}
\label{sec:tables}

Here we have listed all topological materials identified in our scanning of the database, with different types of materials given in different tables. We note that the definition of indicators in Tables~\ref{tab:GMSM}, \ref{tab:TI}, \ref{tab:TCI} follows Ref \cite{Song2017a,Fang2017}, and the notations of irreducible representations of double space groups are the same as those on the BCS. We make a description for the symbols in the tables below:
\begin{itemize}
	\item In Table \ref{tab:HSPSM}, symbols in the second column represent irreducible representations. For example, in symbol $\overline{C}_{2}\overline{C}_{2}(4)$, $C$ is the name of the high symmetry point, $\overline{C}_{2}$ is the name of the irreducible representation defined in the BCS, the overbar means that this is a double group irreducible representation, and 4 in parenthesis denotes the dimension of the irreducible representation. Two irreducible representations written together indicate that the time-reversal symmetry enforces them to be degenerate with doubled dimension.

	\item In Table \ref{tab:HSLSM}, the second column lists all the high symmetry lines that violate compatibility relationships, which reveal the existance of band crossing. The band crossings can be isolated points, parts of nodal lines and nodal surface, which correspond to Weyl/Dirac semimetals, nodal line semimetals and nodal surface semimetals, respectively.

	\item In Table \ref{tab:GMSM}, \ref{tab:TI}, \ref{tab:TCI}, some indicator groups are different from those in Ref \cite{Song2017a,Fang2017}. For example, indicator group for SG 13 is $\mathbb{Z}_{2}\times\mathbb{Z}_{2}$ in the nsoc-setting in Ref \cite{Fang2017}, but is $\mathbb{Z}_{2}\times\mathbb{Z}_{2}\times\mathbb{Z}_{2}\times\mathbb{Z}_{4}$ in our table, which is the indicator of SG 2 in Ref \cite{Fang2017}. This is because SG 13, which is a supergroup of SG 2, has additional symmetries that annihilate two of the three $\mathbb{Z}_2$ generators, forcing them to vanish. This annihilation of weak generators occur in many other space groups, but it is far from obvious which generators are killed, because it depends on the specific space group considered: some times one of the three must vanish and some times two must be equal, etc.. Therefore, we choose to list all three indices for any centrosymmetric space group, and remind the readers that they are not all independent.
	
\end{itemize}

For topological semimetals, including high-symmetry point semimetals, high-symmetry line semimetals and generic momenta semimetals, the ideal Fermi surfaces only consist of nodal points.
However, in reality these materials are also compensate semimetals, having electron and hole pockets that are non-topological.
Due to the linear crossings at topological nodes, the density of states of topological Fermi surface is expected to be smaller compared with that of a normal Fermi surface.
Hence, topological semimetals are considered ``good'' when the density of states at the Fermi energy is small.
For readers to find good materials more easily, we order the topological semimetals in the following ways:
\begin{itemize}
	\item  In Table \ref{tab:HSPSM}, materials having the same space group and the same partially filled high-symmetry points are ordered, from low to high, by their densities of states at the Fermi energy, and we insert ``\textbar{}\textbar{}'' between materials having density of states lower and higher than $0.5$ eV$^{-1}$ per unit cell.
	\item  In Table \ref{tab:HSLSM}, materials having the same space group and the same high-symmetry lines where band crossings appear are ordered, from low to high, by their densities of states at the Fermi energy, and we insert ``\textbar{}\textbar{}'' in each entry between materials having density of states lower and higher than $0.5$ eV$^{-1}$ per unit cell.
	\item In Table \ref{tab:GMSM}, materials having the same space group and the same symmetry-based indicators are ordered, from low to high, by their densities of states at the Fermi energy, and we insert ``\textbar{}\textbar{}'' between materials having density of states lower and higher than $0.5$ eV$^{-1}$ per unit cell.
\end{itemize}

Ideal topological insulators and topological crystalline insulators have a charge gap, but well-defined band topology only requires a full \emph{direct} gap at each momentum. 
In fact, many materials discovered in this class have vanishing indirect gaps.  
In order to give an idea how ``ideal'' one candidate material is, for each material in Table \ref{tab:TI} and Table \ref{tab:TCI}, we have computed the density of states at the Fermi level (not explicitly shown), and by that information order the materials as follows:
\begin{itemize}
	\item In Table \ref{tab:TI}, materials having the same space group and the same symmetry-based indicators are ordered, from low to high, by their densities of states at the Fermi energy, and we insert ``\textbar{}\textbar{}'' between materials having zero density of states and the rest.
	\item In Table \ref{tab:TCI}, materials having the same space group and the same symmetry-based indicators are ordered, from low to high, by their densities of states at the Fermi energy, and we insert ``\textbar{}\textbar{}'' between materials having zero density of states and the rest.
\end{itemize}

In the Tables, if one material has been predicted as topological in a previous publication, the paper is cited immediately after the material formula. Also, if a material formula corresponds to more than one crystal structures in the same space group, we append its ICSD number after it for clarity. Some partially filled $d$- or $f$-shell elements are colored in blue and red respectively in the tables below, as mentioned in Sec. IV A.

\twocolumngrid

\clearpage
\newpage

\newpage
\begin{footnotesize}
\begin{centering}
\begin{longtable*}{>{\centering}m{1.4cm}|>{\centering}m{4cm}|>{\centering}m{10cm} } 
\caption{List of High Symmetry Point Semimetal  \label{tab:HSPSM}} \tabularnewline
\hline
\hline
\hline
\multicolumn{3}{c}{\textbf{List of High Symmetry Point Semimetal }} \tabularnewline
\hline
\textbf{SOC or NSOC} & \textbf{Degenerate Irreducible Representations}  &\textbf{High Symmetry Point Semimetals} \tabularnewline
\hline
\endfirsthead
\multicolumn{3}{c} {{\bfseries \tablename\ \thetable{} -- continued}} \tabularnewline
\hline
\textbf{SOC or NSOC} & \textbf{Degenerate Irreducible Representations}  &\textbf{High Symmetry Point Semimetals} \tabularnewline
\hline
\endhead
\hline \multicolumn{3}{r}{{Continued on next page}} \tabularnewline
\endfoot
\hline \hline
\endlastfoot
\hline
\multicolumn{3}{c}{Space Group 7: $Pc$} \tabularnewline  
\hline
\multirow{1}{*}{NSOC} &  $\rm{A_{1}}$$\rm{A_{2}}$(2), $\rm{B_{1}}$$\rm{B_{2}}$(2), $\rm{D_{1}}$$\rm{D_{2}}$(2), $\rm{E_{1}}$$\rm{E_{2}}$(2)  & 
 || $\rm{{\color{blue}Re}N_{2}}$, $\rm{Ga_{9}{\color{blue}Rh}_{2}}$, $\rm{Ga_{9}{\color{blue}Ir}_{2}}$, $\rm{La_{14}C_{14}Cl_{9}}$ \tabularnewline   
\hline
\hline
\multicolumn{3}{c}{Space Group 9: $Cc$} \tabularnewline  
\hline
\multirow{1}{*}{NSOC} &  $\rm{A_{1}}$$\rm{A_{2}}$(2), $\rm{M_{1}}$$\rm{M_{2}}$(2)  & 
 || $\rm{{\color{red}Ce}{\color{blue}Co}C_{2}}$ \tabularnewline   
\hline
\hline
\multicolumn{3}{c}{Space Group 11: $P2_1/m$} \tabularnewline  
\hline
\multirow{1}{*}{NSOC} &  $\rm{C_{1}}$(2), $\rm{D_{1}}$(2), $\rm{E_{1}}$(2), $\rm{Z_{1}}$(2)  & 
$\rm{TlCuHSeO_{5}}$,  || $\rm{LiBe}$, $\rm{H_{3}W}$$\rm{(ICSD:247596)}$, $\rm{H_{3}W}$$\rm{(ICSD:247600)}$, $\rm{{\color{blue}Re}N_{2}}$, $\rm{TaS_{3}}$, $\rm{Li_{7}Sn_{3}}$, $\rm{ZrCu}$, $\rm{Sb(MoS)_{2}}$, $\rm{LaSi_{2}{\color{blue}Ru}}$, $\rm{NbSe_{3}}$, $\rm{{\color{red}Nd}Si_{2}{\color{blue}Ru}}$, $\rm{Ta{\color{blue}Ni}_{2}Te_{3}}$, $\rm{{\color{red}Pr}Si_{2}{\color{blue}Ru}}$, $\rm{Zr{\color{blue}Rh}}$, $\rm{Hf_{2}{\color{blue}Ni}P}$, $\rm{Zr_{2}{\color{blue}Ni}P}$, $\rm{Y_{3}{\color{blue}Ir}I_{3}}$, $\rm{KYbSiS_{4}}$, $\rm{Bi_{9}I_{2}}$, $\rm{Y_{3}(Ge_{3}Pt_{2})_{2}}$, $\rm{La_{3}BCCl_{3}}$, $\rm{CaCu}$, $\rm{Hf{\color{blue}Ir}}$, $\rm{{\color{red}Sm}_{3}Ti_{3}(SeO_{4})_{2}}$, $\rm{Rb_{9}O_{2}}$, $\rm{YbP_{5}}$, $\rm{Ta{\color{blue}Fe}Te_{3}}$, $\rm{HgSBr}$, $\rm{Cu_{6}AgBi_{12}PbS_{22}}$, $\rm{La_{2}{\color{blue}Ni}_{12}P_{5}}$, $\rm{Ta_{3}Te_{14}Pd_{3}}$, $\rm{Ca_{4}P_{7}{\color{blue}Ir}_{8}}$, $\rm{Ca_{4}As_{7}{\color{blue}Ir}_{8}}$, $\rm{Zr{\color{blue}Cr}_{5}P_{3}}$, $\rm{{\color{red}U}{\color{blue}Cr}_{5}P_{3}}$ \tabularnewline
\hline
\end{longtable*}
\end{centering}
\end{footnotesize}
\newpage
\begin{footnotesize}
\begin{centering}
\begin{longtable*}{>{\centering}m{1.4cm}|>{\centering}m{4cm}|>{\centering}m{10cm} } 
\caption{List of High Symmetry Line Semimetal  \label{tab:HSLSM}} \tabularnewline
\hline
\hline
\hline
\multicolumn{3}{c}{\textbf{List of High Symmetry Line Semimetal }} \tabularnewline
\hline
\textbf{SOC or NSOC} & \textbf{HSL with Band Crossing}  &\textbf{High Symmetry Line Semimetals} \tabularnewline
\hline
\endfirsthead
\multicolumn{3}{c} {{\bfseries \tablename\ \thetable{} -- continued}} \tabularnewline
\hline
\textbf{SOC or NSOC} & \textbf{HSL with Band Crossing}  &\textbf{High Symmetry Line Semimetals} \tabularnewline
\hline
\endhead
\hline \multicolumn{3}{r}{{Continued on next page}} \tabularnewline
\endfoot
\hline \hline
\endlastfoot
\hline
\multicolumn{3}{c}{Space Group 4: $P2_1$} \tabularnewline 
\hline
\multirow{3}{*}{NSOC} &  E-A, D-B   & 
 || $\rm{KNa_{2}AlH_{6}}$ \tabularnewline  
\cline{2-3}
 &E-A   & 
 || $\rm{KCuPO_{4}}$ \tabularnewline  
\cline{2-3}
 &Z-$\Gamma$   & 
 || $\rm{Ag_{2}S}$$\rm{(ICSD:98454)}$ \tabularnewline   
\hline
\hline
\multicolumn{3}{c}{Space Group 5: $C2$} \tabularnewline 
\hline
\multirow{2}{*}{NSOC} &  M-A, Y-$\Gamma$   & 
 || $\rm{KSe_{3}}$, $\rm{Sr_{10}Cu_{5}Bi_{10}O_{29}}$ \tabularnewline  
\cline{2-3}
 &Y-$\Gamma$   & 
$\rm{Na_{5}LiN_{2}}$,  || $\rm{Tl{\color{blue}V}_{5}S_{8}}$, $\rm{RbCuCl_{3}}$, $\rm{{\color{blue}Sc}_{3}{\color{blue}Re}_{2}Si_{3}}$ \tabularnewline   
\hline
\hline
\multicolumn{3}{c}{Space Group 6: $Pm$} \tabularnewline
\hline
\end{longtable*}
\end{centering}
\end{footnotesize}
\newpage
\begin{footnotesize}
\begin{centering}
\begin{longtable*}{>{\centering}m{1.5cm}|>{\centering}m{14cm} } 
\caption{List of Generic Momenta Semimetals \label{tab:GMSM}} \tabularnewline
\hline
\hline
\hline
\multicolumn{2}{c}{\textbf{List of Generic Momenta Semimetals}} \tabularnewline
\hline
\endfirsthead
\multicolumn{2}{c} {{\bfseries \tablename\ \thetable{} -- continued}} \tabularnewline
\hline
\textbf{SI} & \textbf{Generic Momenta Semimetals} \tabularnewline
\hline
\endhead
\hline \multicolumn{2}{r}{{Continued on next page}} \tabularnewline
\endfoot
\hline \hline
\endlastfoot
\hline
\multicolumn{2}{c}{Space Group 2: $P\overline{1}$} \tabularnewline 
\hline
$\mathbb{Z}_{2,2,2,4}$ & Materials \tabularnewline 
\hline
$0001$&  
 || $\rm{{\color{blue}Tc}_{2}As_{3}}$, $\rm{Hf_{8}{\color{blue}Ni}_{21}}$, $\rm{Ti_{5}O_{9}}$ \tabularnewline 
\cline{1-2}
${\color{blue}0002}$&  
 || $\rm{NaCuF_{3}}$, $\rm{In_{3}Cu_{7}}$ \tabularnewline 
\cline{1-2}
$0003$&  
$\rm{Ag_{2}F_{5}}$\cite{mp-542298_mp-568758_mp-130_mp-9027_mp-1008682_mp-1008866_mp-1009020_mp-1008807_mp-28447},  || $\rm{Ba(Mo_{3}S_{4})_{2}}$, $\rm{Sr(Mo_{3}S_{4})_{2}}$, $\rm{CsHg}$, $\rm{{\color{red}Er}_{4}(Al_{8}Pt_{3})_{3}}$, $\rm{Ti_{4}O_{7}}$ \tabularnewline 
\cline{1-2}
$0010$&  
 || $\rm{SbS}$, $\rm{CuMoO_{4}}$, $\rm{K_{3}H{\color{blue}Rh}C_{5}N_{5}O}$ \tabularnewline 
\cline{1-2}
$0011$&  
 || $\rm{Ti{\color{blue}Ni}}$, $\rm{Na{\color{blue}V}_{2}P_{2}(H_{4}O_{7})_{2}}$, $\rm{CuAg_{5}As_{3}O_{11}}$ \tabularnewline 
\cline{1-2}
$0012$&  
 || $\rm{SrAs_{3}}$\cite{mp-9122_mp-1900_mp-11108_mp-15325_mp-1079945}, $\rm{Ta_{6}S}$ \tabularnewline
\hline
\end{longtable*}
\end{centering}
\end{footnotesize}
\newpage
\begin{footnotesize}
\begin{centering}
\begin{longtable*}{>{\centering}m{1.5cm}|>{\centering}m{14cm} } 
\caption{List of Topological Insulators \label{tab:TI}} \tabularnewline
\hline
\hline
\hline
\multicolumn{2}{c}{\textbf{List of Topological Insulators}} \tabularnewline
\hline
\endfirsthead
\multicolumn{2}{c} {{\bfseries \tablename\ \thetable{} -- continued}} \tabularnewline
\hline
\textbf{SI} & \textbf{Topological Insulators} \tabularnewline
\hline
\endhead
\hline \multicolumn{2}{r}{{Continued on next page}} \tabularnewline
\endfoot
\hline \hline
\endlastfoot
\hline
\multicolumn{2}{c}{Space Group 2: $P\overline{1}$} \tabularnewline 
\hline
$\mathbb{Z}_{2,2,2,4}$ & Materials \tabularnewline 
\hline
0001 &  
 || $\rm{{\color{blue}Tc}_{2}As_{3}}$, $\rm{Hf_{8}{\color{blue}Ni}_{21}}$, $\rm{Ti_{5}O_{9}}$ \tabularnewline 
\cline{1-2}
0003 &  
 || $\rm{Ag_{2}F_{5}}$\cite{mp-542298_mp-568758_mp-130_mp-9027_mp-1008682_mp-1008866_mp-1009020_mp-1008807_mp-28447}, $\rm{K_{11}(In_{2}Bi_{3})_{3}}$, $\rm{Ba(Mo_{3}S_{4})_{2}}$, $\rm{Sr(Mo_{3}S_{4})_{2}}$, $\rm{CsHg}$, $\rm{{\color{red}Er}_{4}(Al_{8}Pt_{3})_{3}}$, $\rm{Ti_{4}O_{7}}$ \tabularnewline 
\cline{1-2}
0011 &  
 || $\rm{Ti{\color{blue}Ni}}$, $\rm{Na{\color{blue}V}_{2}P_{2}(H_{4}O_{7})_{2}}$ \tabularnewline 
\cline{1-2}
0101 &  
$\rm{CaP_{3}}$\cite{mp-9122_mp-1900_mp-11108_mp-15325_mp-1079945},  || $\rm{Bi_{3}(TeCl_{5})_{2}}$, $\rm{Yb_{8}Tl_{3}}$ \tabularnewline 
\cline{1-2}
0103 &  
 || $\rm{{\color{red}Pr}_{8}(BCl)_{7}}$, $\rm{NaYb(PS_{3})_{2}}$ \tabularnewline 
\cline{1-2}
0111 &  
 || $\rm{NbCuO_{3}F}$ \tabularnewline
\hline
\end{longtable*}
\end{centering}
\end{footnotesize}
\newpage
\begin{footnotesize}
\begin{centering}
\begin{longtable*}{>{\centering}m{1.5cm}|>{\centering}m{14cm} } 
\caption{List of Topological Crystalline Insulators \label{tab:TCI}} \tabularnewline
\hline
\hline
\hline
\multicolumn{2}{c}{\textbf{List of Topological Crystalline Insulators}} \tabularnewline
\hline
\endfirsthead
\multicolumn{2}{c} {{\bfseries \tablename\ \thetable{} -- continued}} \tabularnewline
\hline
\textbf{SI} & \textbf{Topological Crystalline Insulators} \tabularnewline
\hline
\endhead
\hline \multicolumn{2}{r}{{Continued on next page}} \tabularnewline
\endfoot
\hline \hline
\endlastfoot
\hline
\multicolumn{2}{c}{Space Group 2: $P\overline{1}$} \tabularnewline 
\hline
$\mathbb{Z}_{2,2,2,4}$ & Materials \tabularnewline 
\hline
$0002$ &  
 || $\rm{NaCuF_{3}}$, $\rm{In_{3}Cu_{7}}$ \tabularnewline 
\cline{1-2}
$0010$ &  
 || $\rm{SbS}$, $\rm{CuMoO_{4}}$ \tabularnewline 
\cline{1-2}
$0012$ &  
 || $\rm{SrAs_{3}}$\cite{mp-9122_mp-1900_mp-11108_mp-15325_mp-1079945}, $\rm{Ta_{6}S}$ \tabularnewline 
\cline{1-2}
$0100$ &  
 || $\rm{{\color{blue}Ni}_{13}Sn_{8}P_{3}}$ \tabularnewline 
\cline{1-2}
$0102$ &  
 || $\rm{KHg}$, $\rm{Zr_{8}{\color{blue}Ni}_{21}}$, $\rm{CuH_{18}C_{4}N_{10}(ClO)_{2}}$ \tabularnewline 
\cline{1-2}
$0110$ &  
 || $\rm{Pt_{2}PbO_{4}}$, $\rm{CsCuP_{3}H_{8}O_{13}}$, $\rm{{\color{blue}Ni}_{10}Sn_{5}P_{3}}$, $\rm{SrTi_{11}O_{20}}$ \tabularnewline
\hline
\end{longtable*}
\end{centering}
\end{footnotesize}

\clearpage

\end{document}